# Nanoscale Resistive switching in electrodeposited MOF Prussian blue analogs driven by K-ion intercalation probed by C-AFM

L. B. Avila[1], O. de Leuze[1], M. Pohlitz[2], M.A. Villena[3], Ramon Torres-Cavanillas[4], C. Ducarme[1], A. Lopes Temporao[1], T. G. Coppée[1], Anatole Moureaux[1], S. Arib[1], Eugenio Coronado[4], C. K. Müller[2], J. B. Roldán[3], B. Hackens[1], F. Abreu Araujo[1]

[1]Institute of Condensed Matter and Nanosciences (ICMN), Université catholique de Louvain (UCLouvain), Louvain-la-Neuve B-1348, Belgium.

[2]Faculty of Physical Engineering/Computer Sciences, University of Applied Sciences Zwickau, 08056 Zwickau, Germany.

[3]Departamento de Electrónica y Tecnología de Computadores, Universidad de Granada, Facultad de Ciencias, Avd. Fuentenueva s/n, 18071 Granada, Spain.

[4]Instituto de Ciencia Molecular, Universitat de Valencia, Valencia, 46980 Spain

Corresponding author email: lindiomar.borges@uclouvain.be; flavio.abreuaraujo@uclouvain.be

## ABSTRACT

K-ion intercalation in Prussian blue analogues (PBAs) is a well-established charge storage mechanism in potassium-ion batteries; here, we demonstrate that this process also responsible for nanoscale resistive switching in PBA-based material. Using conductive atomic force microscopy (C-AFM), we directly visualize and electrically control reversible conductance modulation within sub-100 nm volumes, linking localized $K^+$ redistribution to $Fe^{2+}/Fe^{3+}$ redox reconfiguration. The switching is polarity-selective, with Prussian white (PW) and Prussian blue (PB) both exhibiting unidirectional resistive switching (URS) at opposite bias polarities, consistent with oxidation- and reduction-driven transport, respectively. These processes are governed by electrically confined ion–electron coupling, where K-ion motion modulates small-polaron hopping within the PBA framework. The switching dynamics are strongly dependent on ionic mobility, with PW sustaining operation up to 200 V/s and PB up to 50 V/s, highlighting the critical role of K-ion concentration in enabling fast redox kinetics. This nanoscale switching mechanism, combined with spatial confinement below 100 nm, enables high-density device integration without interference between them. Moreover, PBAs provide a chemically versatile, earth-abundant platform with tunable ionic and electronic properties, fabricated via a single-step, aqueous, room-temperature process compatible with CMOS technology. This solution-processable approach enables low-temperature fabrication, scalability to large-area substrates, and integration into diverse device architectures. The use of simple salt precursors ensures low-cost manufacturing, while environmentally friendly synthesis and non-toxic, non-contaminating disposal enhance overall sustainability.

INTRODUCTION

With the rapid expansion of big data and artificial intelligence, the demand for computing technologies capable of delivering high processing speed and low energy consumption has grown dramatically, especially for tasks such as pattern recognition, real-time image processing, and autonomous decision-making [1–3]. However, modern computing still relies predominantly on the traditional von Neumann architecture, in which memory and processing units are physically separated. This structural distinction imposes a severe bottleneck on data throughput and leads to substantial energy losses during frequent memory access [4–6]. In contrast, the human brain performs computation in a massively parallel fashion through densely interconnected networks of neurons and synapses, achieving remarkable efficiency with only a few tens of watts of power consumption [7,8].

Memristive devices whose conductance can encode a synaptic weight, offer a promising route toward hardware systems that emulate synaptic plasticity and neural information processing [9–12]. Multiple switching mechanisms have been employed to reproduce biological functionality, including conductive filament formation, phase transitions, magnetization switching, and ion migration [13–16]. Among these, ion migration-based memristors are especially attractive because their operation closely mimics the ionic dynamics of biological synapses, including $Ca^{2+}$driven plasticity, and because they naturally support analog state evolution [17,18]. These ion-driven effects have inspired the development of battery-like synapses, where electrolytes and electrochemical reactions are used to modulate device conductance at extremely low power [19,20]. However, many of such devices still exhibit millisecond-scale switching times, posing challenges for high-speed neuromorphic processing. Given that ion transport often dictates switching speed, regulation of intercalation kinetics has emerged as a critical pathway for enabling next generation, high-speed, energy-efficient synaptic devices [21].

Ion intercalation is known to induce pronounced structural and electronic modifications in layered and framework materials [22]. For example, multilayer $2H - MoS2$, with its large interlayer spacing of $0.62\ nm$, offers abundant storage sites and rapid diffusion channels for $Li^+$ ions, enabling reversible conductance modulation under an applied field [22–24]. Porous electrochemical solids represent another particularly attractive class of materials in this context. In particular, metal–organic frameworks and coordination polymers offer exceptional opportunities because their composition, redox chemistry, and host–guest interactions can be tuned synthetically, allowing the electrochemical window of the lattice, and thus the ion-migration-driven memristive response, to be engineered by design [25-30]. However, many such systems still suffer from limited cycling stability and spatially inhomogeneous ion migration, which hinder reproducible switching and long-term device operation.

In this context, Prussian blue (PB) and its analogues (PBAs), among the earliest known coordination polymers, have emerged as especially appealing candidates for iontronic switching applications. Their attractiveness stems from a rigid open framework with well-defined cavities for alkali-ion insertion and stabilization, combined with a redox-active transition-metal lattice that can sustain both filamentary and ion-migration-driven memristive switching over hundreds to thousands of cycles without significant structural fatigue [31,32]. PBAs can be described by the general formula:

$$AQ[M(CN)_6]$$

where ($A = Li, K, Na$; $Q = Fe, Co, Ni, Mn, Cu$; $M = Fe, Mn, Co, etc.$), in which alkali-metal ions occupy spacious cubic cavities within a rigid $Fe - C \equiv N - Fe$ network [33,34]. In these materials, alkali-metal ions occupy spacious cubic cavities within a face-centered cubic cyanometallate network composed of transition-metal octahedra linked by cyanide ligands. This intrinsic architecture gives rise to three-dimensional ion-diffusion pathways, strong coupling between ion distribution and $Fe^{2+}$/$Fe^{3+}$ redox states, and electronically tunable mixed-valence transport [35]. Consequently, PBAs have recently gained attention as versatile candidates for ion-intercalation memristors and neuromorphic devices.

Despite this promise, mechanistic insight into the local switching processes of PB/PBA-based memristors remains limited. In particular, few studies have probed memristive behavior with sufficient spatial resolution to directly resolve the ionic dynamics that underlie it. As a consequence, the relationship between local ion migration, redox redistribution, and resistive switching remains poorly understood, hindering the rational design of PBA-based iontronic devices.

Here, we address this challenge by locally investigating the memristive response of two representative end members of the PBA family, Prussian white (PW) and Prussian blue (PB), which differ primarily in their $K^+$ content and Fe oxidation-state distribution [36,40]. This chemically defined contrast provides an ideal platform for elucidating how alkali-ion content governs nanoscale resistive switching. By combining conductive atomic force microscopy (C-AFM), finite-element simulations, and in situ Raman spectroscopy, we correlate the local memristive response of electrodeposited PW and PB thin films with their intrinsic $K^+$ concentration and redox chemistry. We show that these differences directly determine the switching polarity, with both PW and PB exhibiting URS at opposite bias polarities, and demonstrate that the resistive switching is governed by a nanoscale, electrically confined ion–electron coupling mechanism involving K-ion redistribution and redox-modulated small-polaron transport beneath the probe [41,42]. Beyond clarifying the local switching mechanism, our results establish PBAs as a chemically tunable and scalable materials platform for low-dimensional memristive devices and neuromorphic computing. Their practical appeal is further strengthened by single-step aqueous electrodeposition at room temperature from earth-abundant precursors, providing a low-cost, scalable, and CMOS-compatible route to device fabrication.

RESULTS

In Fig. 1, elemental maps of $Fe$ and $K$ obtained by energy dispersive X-ray spectroscopy (EDS) provide direct evidence of the precise compositional control achieved in electrodeposited PBA thin films. Samples deposited at potentials between $0.1$ and $0.25$ V can be clearly identified as Prussian White (PW), whereas films deposited at potentials above $0.28$ Vcorrespond to the Prussian Blue (PB) phase. The EDS maps reveal a well-defined and systematic transition from PW to PB with increasing electrodeposition potential. Films grown at lower potentials exhibit a distinctly K-rich composition, indicative of enhanced incorporation of interstitial $K+$ ions required to charge-compensate the reduced framework dominated by $Fe^{2+}/Fe^{2+}$ pairs. In contrast, films deposited at higher potentials show a markedly lower K content, consistent with the mixed-valence $Fe^{2+}/Fe^{3+}$ configuration characteristic of the oxidized PB phase. The corresponding chemical formulas of the two phases are presented in Fig. 1.

This potential-dependent modulation of alkaline ion incorporation demonstrates that the electrodeposition process provides precise and reproducible control over both the redox state of iron and the interstitial $K^+$ population. This control directly governs the local electrostatic landscape, ionic shielding, and electronic charge distribution in the parameters of the films that play a central role in determining charge transport and resistive switching behavior in PBA-based devices. For the present study, we selected two samples with compositions well separated from the phase transition regime to ensure structural correlations. PW deposited at $0.1\,V$, as shown in Fig. 1, incorporates a substantially larger fraction of interstitial $K^+ \approx 50\%$, while PB deposited at $0.3\,V$ contains a reduced $K^+$ content of approximately $\approx 30\%$. In addition to the composition, the samples are easily distinguishable by their coloration, as shown in the photos in Fig. 1. This well-defined and adjustable variation in alkaline ion charge is a striking characteristic of electrodeposited PBA thin films and provides a robust platform for systematically investigating the influence of ionic content on electrical functionality. Additional morphological information is provided in Figures S1 and S2.

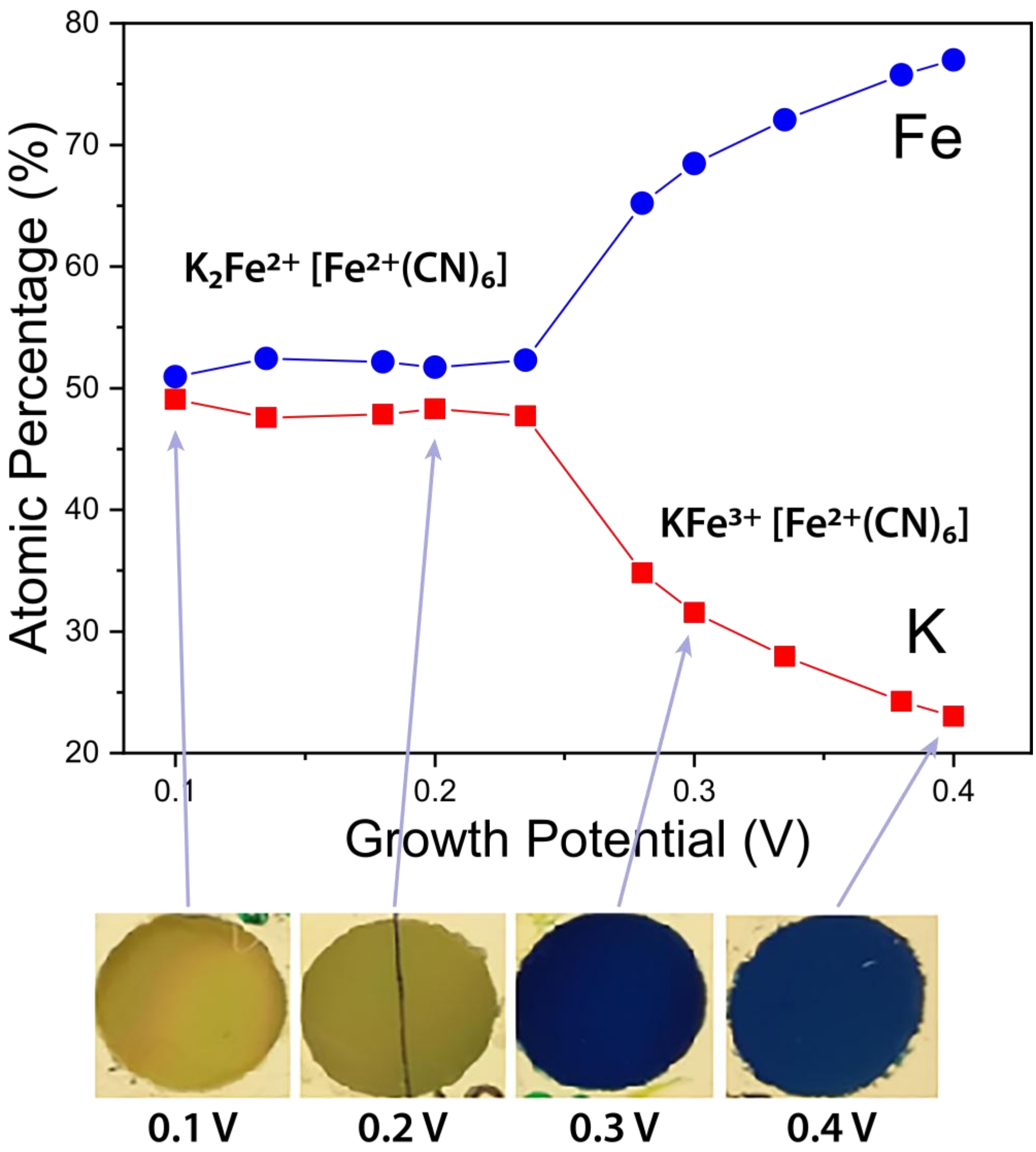

Figure 1. Atomic percentages derived from energy-dispersive X-ray spectroscopy (EDS) elemental maps of Fe and K illustrating controlled evolution of composition from PW to PB. As the electrodeposition potential increases, the materials exhibit a clear transition from K-rich PW to progressively K-deficient PB.

C-AFM measurements were performed to probe the nanoscale resistive switching behavior of the two PBAs phases. Current maps acquired over $1 \times 1\ \mu m^2$ areas at a bias of $1\ V$ (Figures 2b and 2e) reveal clear conductance differences between the two phases. This contrast in conductivity is directly linked to the conduction mechanism of these coordination polymers, which is governed by polaron-assisted transport arising from mixed $Fe^{2+}/Fe^{3+}$ metal canters [43]. In PW, the reduced fraction of $Fe^{3+}$, at the expense of an increased $Fe^{2+}$ content, suppresses this mixed-valence pathway, resulting in a lower overall conductivity. This behavior is consistent with the theoretically predicted widening of the electronic bandgap in PW [44]. The blue crosses overlaid on the current maps in Figures 2b and 2e mark the precise nanoscale locations at which individual I–V measurements were acquired using the C-AFM probe. These marked points correspond directly to the I–V curves presented in Figures 2c and 2f, where the electrical response of each selected site is shown in detail. The resulting I–V characteristics exhibit well-defined SET and RESET transitions, clearly revealing the transition between the low-resistance (LRS) and high-resistance (HRS) states for both PW and PB. The measurements were carried out using a positive-bias sweep from $0$ to $10\ V$ at a constant ramp rate of $0.1\ V/s$ (same protocol applied uniformly across all experiments), this controlled sweep condition enables consistent probing of the switching polarity and reliable quantification of the resistive-state modulation. Additional nanoscale mapping details are provided in Figures S3 and S4, including C-AFM topography maps acquired prior to I–V measurements at varying point spacings

From the perspective of the existing literature, the conductivity of PB is primarily governed by polaron hopping. The memristive behavior of PB films can be rationalized in a manner analogous to that of light-emitting electrochemical cells [45]. PB contains mobile counterions that enable reversible oxidation and reduction of the Fe centers, as $K^+$ ions can migrate through the framework cavities to locally compensate charge. Under positive tip bias, electrons are extracted from the material and collected at the tip, while the substrate supplies electrons, promoting local oxidation processes associated with $Fe^{2+} \rightarrow Fe^{3+}$ conversión. Above a characteristic threshold voltage (approximately $2 - 4\ V$ for PB and around $7\ V$ for PW, Fig. 2) the $I - V$ characteristics exhibit rectifying behavior. Below threshold, current is injection-limited (contact barrier– limited), leading to low residual conduction. We extract the exact mechanism via the data treatment of the current vs voltage data. Once the threshold is exceeded, efficient electron and/or hole injection occurs, leading to an exponential increase in current. In this context, this injection of charge carriers also alters the oxidation state of the $Fe$ centers, thereby generating the memristive response, which persists after the applied bias is removed.

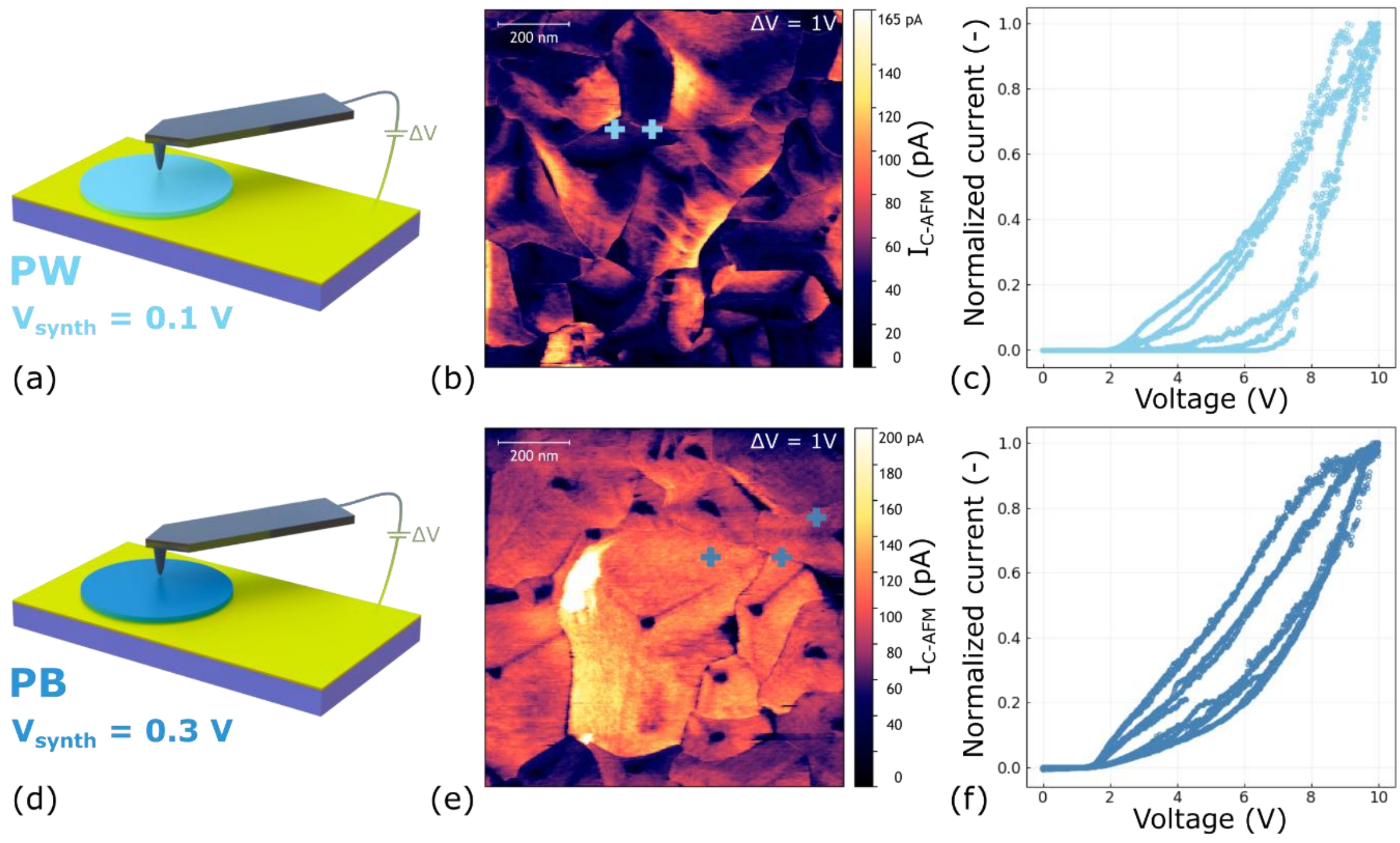

Figure 2. C-AFM measurements on PBA samples. (a), (d): illustrations of the measurement configuration on Prussian white (PW) and Prussian blue (PB) samples, respectively. (b) Current map obtained with a 1 × 1 µm C-AFM scan on (b) PW and (e) PB with a 1 V bias voltage. Multiple I-V sweeps (c) PW and (f) PB samples acquired at different locations of the C-AFM scans, shown by the markers in (b) and (e).

To obtain a more precise evaluation of the RS behavior in the two PBA phases, local I-V scans by C-AFM were performed in a hysteresis range of -10 to +10 V, with a scan rate of 1 V/s (Fig. 3). The I-V characteristics of the PW film (Fig. 3a) exhibit clear URS, where the SET and RESET transitions occur under the same voltage polarity. This nanoscale URS response is fully consistent with previous macroscopic measurements obtained using conventional upper and lower electrode configurations, as reported for PW-based devices in the literature [36, 37, 46, 47], with a predominance of the current response in the negative polarity. In contrast, the PB film exhibits URS behavior (Fig. 3b), with a smaller electrical response and a predominance of the current response now in the positive polarity. C-AFM I–V measurements demonstrate that both PBA phases, PW and PB retain intrinsic nanoscale resistive switching, with both systems exhibiting URS. In each case, the switching is confined to a single bias polarity, although the preferred polarity is opposite for PW and PB. This polarity-dependent behavior is consistent with distinct redox-driven conduction mechanisms in the two phases. Upon closer examination, the memristive response in PW is attributed to oxidation-induced conductivity enhancement, whereas in PB it is associated with reduction. This interpretation is consistent with the redox states of the materials: PW, as the most reduced phase, has limited capacity for further reduction, such that slight oxidation increases charge carrier density and activates polaron hopping pathways. In contrast, PB, which is intrinsically mixed-valent and may deviate from ideal stoichiometry toward a partially over-oxidized state, exhibits enhanced conductivity upon mild reduction. In both cases, the switching behavior can therefore be rationalized in terms of redox-modulated small-polaron transport within the PBA framework.

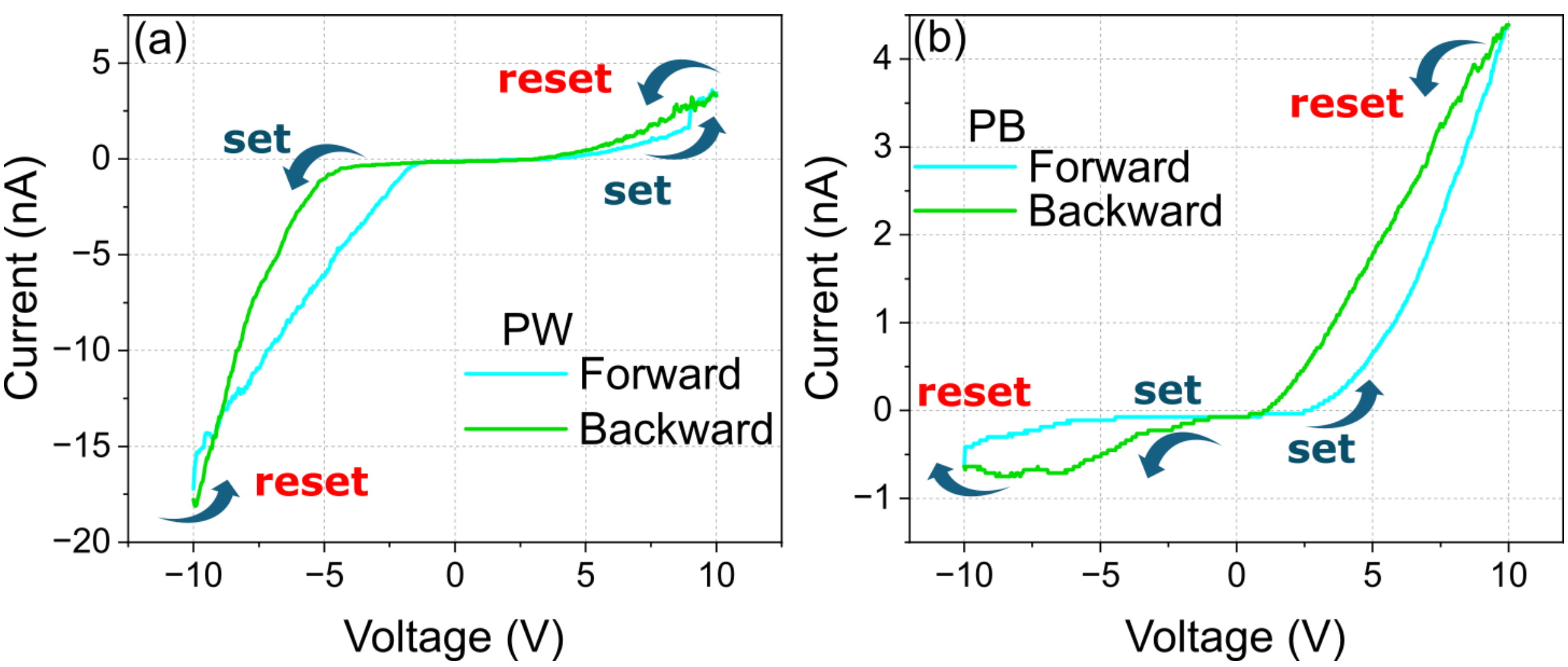

Figure 3. Current–voltage characteristics measured by C-C-AFM. (a) PW film exhibiting unidirectional resistive switching at a single bias polarity. (b) PB film also showing URS, with switching occurring at the opposite polarity to PW.

To assess the dynamical limits of the RS mechanism, voltage-ramp measurements were performed at a fixed location, and varying the sweep rate. Although C-AFM offers high spatial resolution, it is not suitable for endurance cycling. Repeated measurements can induce tip drift, contact degradation, and local surface damage, while the measured current is also sensitive to time-dependent effects and ambient conditions such as relative humidity [48,49]. Consequently, changes observed during prolonged cycling may reflect measurement artifacts rather than intrinsic material behavior, making C-AFM better suited for nanoscale mapping and mechanistic studies than for endurance testing. Therefore, to probe the intrinsic RS behavior of PW and PB surfaces, 10 consecutive C-AFM voltage sweeps were performed from 0 to +10 V at the same spatial location with a scan rate of $0.1V/s$. This voltage range was selected to avoid polarity-induced conditioning effects observed in full bipolar sweeps (−10 to +10 V), which introduce history-dependent variability in the current response, thereby ensuring reproducible measurements and direct comparison between PW and PB under identical conditions.

The corresponding results are shown in Figures S5 and S6 for PW and PB, respectively. The I–V characteristics consistently exhibit the key features of resistive switching, confirming that the effect is reliably captured at the nanoscale. Although cycle-to-cycle variability is observed, consistent with nanoscale measurement conditions. Such variability is expected given the inherent sensitivity of C-AFM measurements to tip–sample contact stability, local surface modifications, and environmental factors as was already mentioned above. Importantly, the persistence of the switching response across repeated sweeps demonstrates the robustness of the RS mechanism in both materials.

After characterizing the switching behavior at a fixed location, we examined the effect of voltage sweep rate on the resistive switching. Measurements were conducted at ten distinct locations on each sample, separated by 500 nm, to account for spatial variability. For the PW films, voltage ramps were applied over a range of sweep rates (0.2, 0.5, 1.0, 2.0, 10, 50, 100, and 200 V/s), shown in Figure S7. Up to 2 V/s, the current levels and switching characteristics remained comparable to those presented in Fig. S6. At higher sweep rates, the resistive switching behavior remained clearly up to 200 V/s, although with reduced current levels. Above this value, the switching characteristics were no longer observed. For the PB films, we observe a similar trend as in Figure S8, but with a lower kinetic threshold: resistive switching persisted only up to 50 V/s, with the switching signature disappearing at higher sweep rates. The RS in both PW and PB can be understood as a redox-driven, ion-assisted modulation of small-polaron transport within the PBA framework, with distinct behavior arising from their different initial redox states and $K^+$ content.

In PW, which is the most reduced phase, the application of an electric field induces local oxidation, generating mixed-valence $Fe^{2+}/Fe^{3+}$ (will be discussed in more detail later in this paper) sites that increase charge carrier density and activate small-polaron hopping. The relatively high concentration of $K^+$ ions (Fig. 1) facilitates rapid ionic redistribution, which supports fast redox equilibration. As a result, PW exhibits efficient and fast switching, maintaining RS behavior even at high scan rates (up to 200 V/s). In contrast, PB is intrinsically mixed-valent but likely partially over-oxidized. Under an applied field, mild reduction restores a more favorable $Fe^{2+}/Fe^{3+}$ balance, enhancing electron hopping and increasing conductivity. However, the lower $K^+$ content in PB limits ionic mobility, slowing down the redox processes required for switching. Consequently, PB displays slower switching kinetics, with RS observable only at lower scan rates (≤50 V/s), and a polarity preference corresponding to reduction.

We showed that the RS effect persists when performing repeated measurements at the same location and that it remains observable across a range of voltage sweep rates. To further demonstrate nanoscale control over the RS behavior in PW and PB films, systematic conductive C-AFM I–V mapping was carried out. These measurements were conducted using the instrument's dedicated control software, which enables precise definition of both the spatial distribution and the number of electrical probing points with nanometer-scale accuracy, allowing controlled and reproducible investigation of the local switching behavior. For each sample, predefined grids with controlled inter-point spacing and scan area were programmed, allowing reproducible and spatially resolved electrical characterization (PW in Fig. 4 and PB in Fig. 5). The exact measurement locations are indicated by crosses in Figures 4 and 5, illustrating the high degree of positional control achieved.

Three mapping configurations were employed for both PW and PB films. First, a $5 \times 5\ \mu m^2$ area was probed using a $10 \times 10$ grid (100 points) with a lateral spacing of $500\ nm$ between adjacent points (PW in Fig. 7a and PB in Fig. 8a), ensuring representative sampling over a relatively large surface region. The mapping area was then reduced to $2.5 \times 2.5\ \mu m^2$, with the inter-point spacing decreased to $200\ nm$, resulting in an $11 \times 11$ array comprising 121 measurement points (PW in Fig. 4b and PB in Fig. 5b). Finally, high-resolution mapping was performed over a $1 \times 1\ \mu m^2$ area using a $10 \times 10$ grid with a spacing of $100\ nm$ (PW in Fig. 4c and PB in Fig. 5c), enabling detailed nanoscale assessment of local RS variability.

At each measurement site, two consecutive $I - V$ sweeps from $0$ to $10\ V$ were recorded to probe the local RS response while limiting cumulative electrical stress on the active layer, all the raw data is available on the GitHub link in the methods section. Although C-AFM is a powerful technique for investigating RS phenomena with nanometer-scale resolution, repeated electrical cycling at a single location can lead to irreversible local modifications, including Joule heating, filament overgrowth, or mechanical degradation of the tip–sample contact. To avoid such artifacts and to preserve the intrinsic switching behavior, each mapping configuration was therefore performed on a fresh, nonoverlapping region of the film surface.

This measurement strategy is consistent with the methodology established by M. Lanza and co-authors [49], who demonstrated that C-AFM is an effective approach for vali dating RS in ultrashort metal–insulator–metal (MIM) cells by emulating nanoscale device geometries. In that work, the authors emphasized that C-AFM is well suited for spatially resolved RS studies and variability analysis. Following these guidelines, our approach prioritizes spatial mapping and statistical assessment of RS behavior over extended cycling at individual sites, ensuring reliable nanoscale characterization.

After completing all I–V measurements on each array, every point was analyzed individually to determine whether RS occurred. A point was classified as exhibiting resistive switching when the ratio between the high-resistance state (HRS) and low-resistance state (LRS) currents exceeded a defined threshold of HRS/LRS $\geq 1.5$ . These ratios were then plotted as heat maps, enabling visualization of the spatial distribution of switching activity across the nanoscale arrays. The performance of the RS effect across each measurement grid was evaluated by analyzing every individual I–V curve and representing the results in color-coded maps. For each measurement site, the ratio between the HRS and LRS currents was extracted and used as a quantitative metric of switching functionality. Points exhibiting an HRS/LRS ratio above a predefined threshold were classified as functional RS cells, whereas points with a very low or near-zero ratio were considered non-functional. These ratios were then visualized as heat maps, allowing for easy assessment of both the spatial distribution and the relative intensity of RS behavior in nanoscale matrices.

In the color scale, non-functional cells are represented by dark or near-zero values, while cells in the blue-to-green range correspond to moderate, yet reliable, HRS/LRS ratios. Higher performance switching sites, characterized by large resistance contrasts, appear in the green-to-yellow region of the map, indicating enhanced RS robustness. As shown in Fig. 4d for PW and Fig. 5d for PB, the $500\ nm$-spaced arrays reveal a high density of functional cells, with PW exhibiting RS at $99\%$ of the measured points and PB at 88%. When the interpoint spacing was reduced to $200nm$, the fraction of active sites remained high, reaching $96\%$ for PW and $95\%$ for PB. At the highest spatial resolution of $100nm$, PW maintained a consistently high yield of $96\%$, whereas PB showed RS at $83\%$ of the probed locations. The color map analysis provides a clear visualization of both the spatial uniformity and the robustness of the resistive switching response. The PW film exhibits a more homogeneous distribution of active switching sites and a larger fraction of locations with high HRS/LRS ratios, represented by the yellow regions in the map. This indicates not only a higher yield of functional cells but also more stable and pronounced resistance contrast across the probed area.

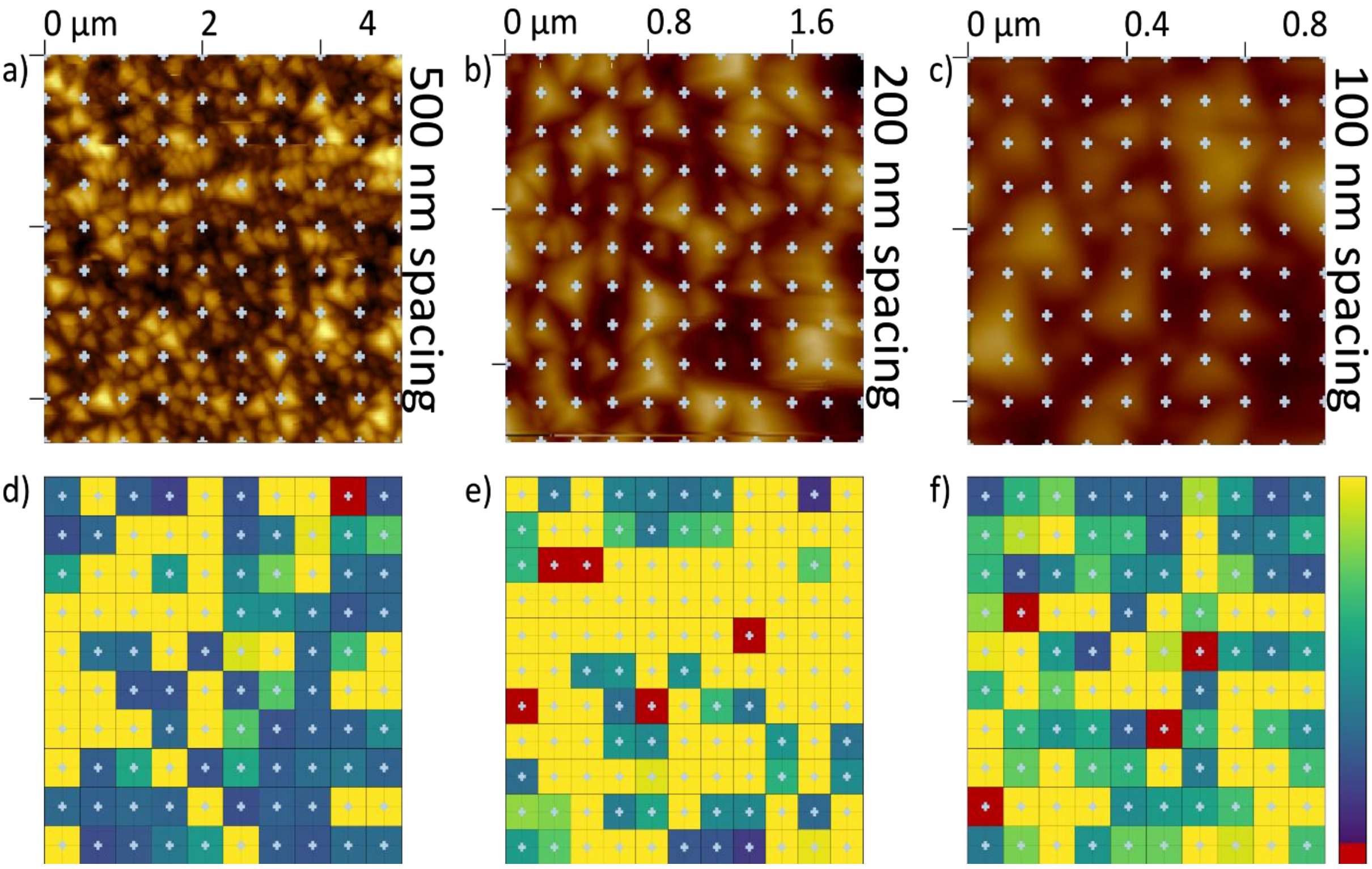

Figure 4. C-AFM map. (a–c) Topography maps of the PW film acquired over areas of $5 \times 5\ \mu m^2$, $2.5 \times 2.5\ \mu m^2$, and $1 \times 1\ \mu m^2$, respectively. The markers (crosses) indicate the precise locations at which $I - V$ curves were recorded, with inter-point spacings of $500\ nm$, $200\ nm$, and $100\ nm$, respectively. (d–f) Color-coded heat maps of the HRS/LRS current ratio extracted from C-AFM I–V measurements on the PW film, corresponding to the measurement grids shown in (a–c). The maps were obtained for inter-point spacings of $500\ nm$ (d), $200\ nm$ (e), and $100\ nm$ (f). Non-functional cells, exhibiting a very low or near-zero HRS/LRS

ratio, are shown in red, whereas functional switching sites appear from blue to green. PW exhibits high RS yields of 99%, 96%, and 96% for the 500 $nm$, 200 $nm$, and 100 $nm$ grids, respectively.

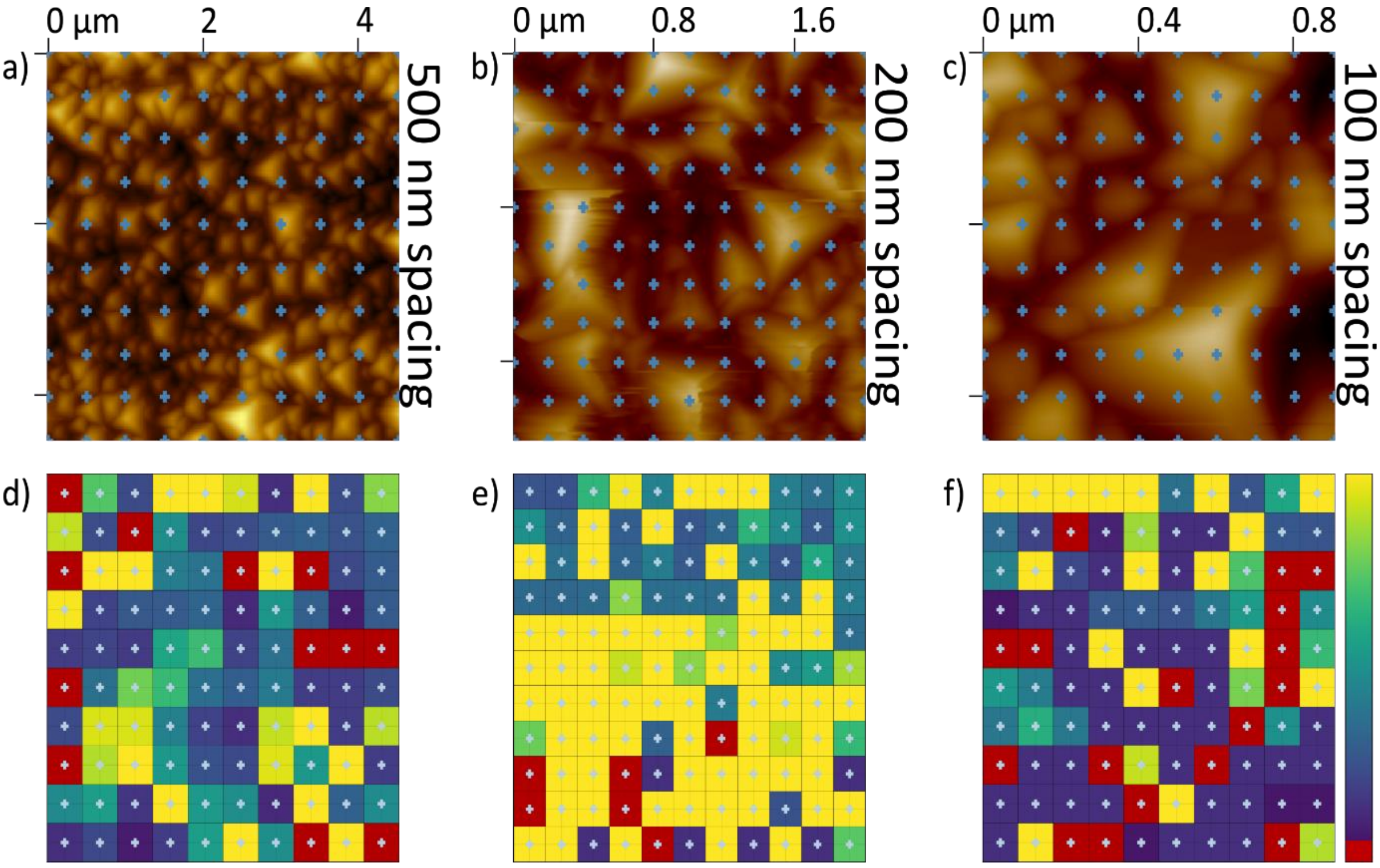

Figure 5. C-AFM map. (a–c) Topography maps of the PB film acquired over areas of $5 \times 5\ \mu m2, 2.5 \times 2.5\ \mu m2, and\ 1 \times 1\ \mu m2$, respectively. The markers (crosses) indicate the precise locations at which I–V curves were recorded, with inter-point spacings of 500 $nm$, 200 $nm$, and 100 $nm$, respectively. (d–f) Color-coded heat maps of the HRS/LRS current ratio extracted from C-AFM I– V measurements on the PB film, corresponding to the measurement grids shown in (a–c). The maps were obtained for inter-point spacings of 500 $nm$ (d), 200 $nm$ (e), and 100 $nm$ (f). Non-functional cells, exhibiting a very low or near-zero HRS/LRS ratio, are shown in red, whereas functional switching sites appear from blue to green. PB exhibits RS yields of 88%, 95%, and 83% for the 500 $nm$, 200 $nm$, and 100 $nm$ grids, respectively.

To obtain a deeper understanding of how individual nanoscale memristors interact within the mapped arrays and to gain a more detailed view of the switching mechanism itself, finite-element simulations were performed to model the tip induced electric field during C-AFM measurements. While such simulations have traditionally been used to interpret C-AFM responses in isotropic dielectric or semiconductor materials [50-52], they have not been applied to PBAs. Here, the model was established considering 25 $nm$ tip radius and micrometer-scale equipotential layers representing the lateral extent of the PW and PB films. The resulting potential distributions (Fig. 6 left, right) show that the electric field is highly localized: the potential penetrates effectively only up to $\sim 60\ nm$ both laterally and vertically from the tip–sample junction. This short penetration depth is significantly smaller than the smallest spacing used in our experimental arrays (100 $nm$), demonstrating that neighboring memristors do not influence one another. Each memristive site, therefore, behaves as an electrically independent nanoscale cell, and measurements performed on adjacent points do not introduce crosstalk or interference.

The simulations also clarify the underlying physical origin of this strong localization. Because PBAs contain interstitial $K^+$ ions, their interaction with the applied electric field is important: $K^+$ experiences a force along the field direction and migrates from regions of lower to higher electric potential. This field-driven redistribution contributes to short-range ionic shielding, further confining the electric field and modifying the local conductivity pathways only within the immediate vicinity of the C-AFM tip. Consequently, the tip–sample contact behaves as an effectively isotropic region, because the electric field is confined to a $\sim 60\ nm$ volume where the cubic $Fe - C \equiv N - Fe$ framework, together with shortrange $K^+$ ionic shielding, generates a locally homogeneous electrostatic environment. This short-range shielding arises from the fact that interstitial $K^+$ ions can undergo only limited displacements within their coordination cavities, enabling them to partially compensate for the applied field but only over nanometric distances. The resulting confinement of the electric perturbation prevents lateral field propagation, ensuring that each memristive site in the array operates independently, without interference even at the smallest spacing of $100\ nm$. This local isotropy is consistent with the intrinsic tridimensional connectivity of the PBA structure. Furthermore, analysis of experimental measurements confirms that there is no dependence between the thickness of the PBA layer and the electrical response of the material (see SI Figure S9). This demonstrates that the switching process is a local process that takes place in the region of penetration of the electric field.

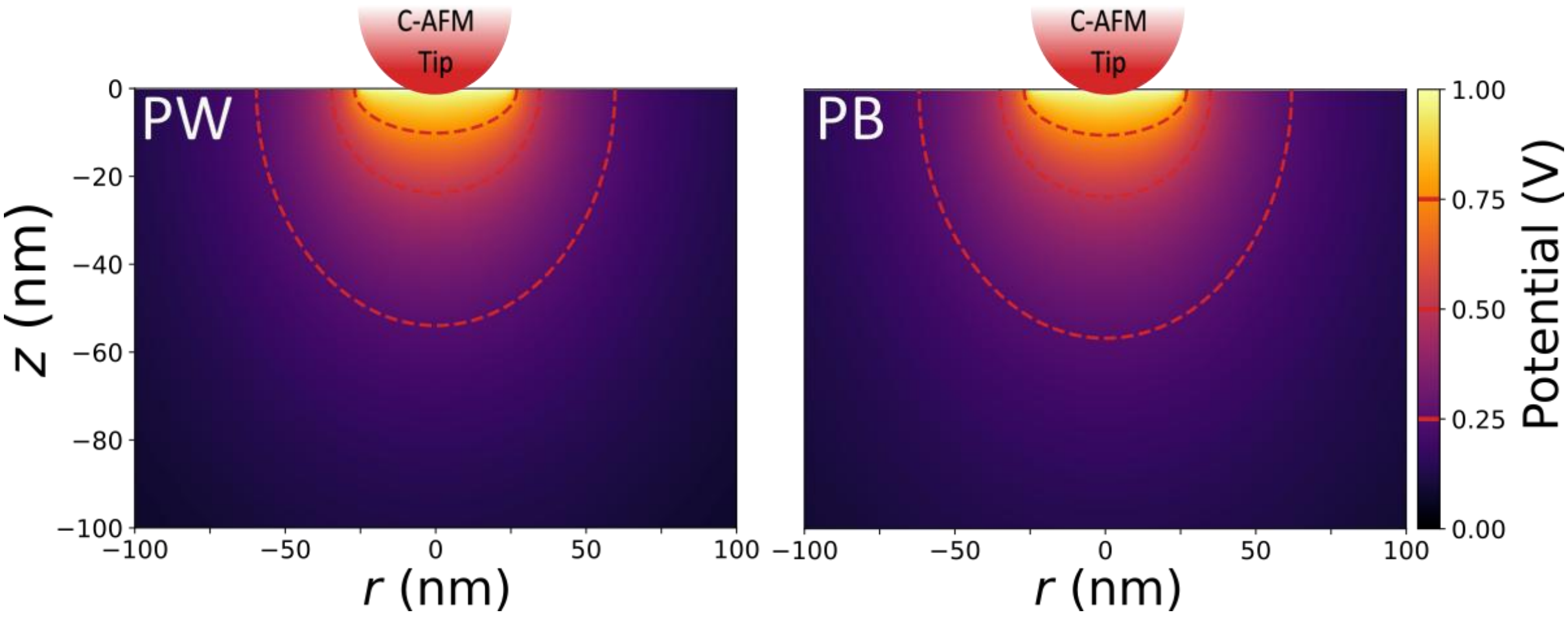

Figure 6. Simulated electrostatic potential distribution below the conductive C-AFM tip for (left) Prussian white and (right) Prussian blue, illustrating the strong nanoscale confinement of the electric field within the PBA framework.

Fig. 7 shows how the electric field generated by the CAFM tip modifies the PW, providing spectroscopic evidence of the switching mechanism being a combination of oxidation/ reduction and $K$-ion intercalation. In situ Raman spectroscopy was performed directly at the site where $I - V$ scans were performed with C-AFM, ensuring a precise correlation between the electrical stimulus and the resulting chemical response at the site. The optical image shows the $5 \times 5$ $\mu m^2$ region previously examined by C-AFM (Fig. 7a). The arrow indicates the position and direction of the scan in which the 50 consecutive spectra were acquired, allowing high-resolution monitoring of the vibrational changes induced by the electric field.

The Raman spectra reveal clear modifications in the PW film when an external voltage sweeps. In the as-prepared state, the film exhibits three principal CN stretching contributions: bands near $\sim$ 2080 and $\sim$ 2120 $cm^{-1}$, attributed to $Fe^{2+} - CN - Fe^{2+}$ linkages and/or uncoordinated $Fe(CN)_6^4$ units, together with a higher-frequency component at $\sim$ $2155 - 2160 cm^{-1}$ associated with $Fe^{2+} - CN - Fe^{3+}$ linkages. These assignments are consistent with reported bulk PW references, which also contain PB-like $Fe^{2+} - CN - Fe^{3+}$ species (presented in Figure S10). Thus, even in the nominally reduced PW state, the film contains a mixed-valence population rather than a purely $Fe^{2+} - CN - Fe^{2+}$ configuration. Upon voltage application, the intensity of the $Fe^{2+} - CN - Fe^{3+}$ band at $\sim$ 2156 $cm^{-1}$ increases at the expense of the lower frequency modes. This redistribution of spectral weight indicates electrically induced oxidation of $Fe^{2+}$ toward $Fe^{3+}$, enhancing PB-like linkages within the PBA lattice

The spatial distribution of this redox modulation is visualized in Fig. 7b through a 2D confocal Raman spectroscopy map centered on the $\sim$ 2156 $cm^{-1}$ band. Brighter regions correspond to areas with a higher fraction of $Fe^{2+} - CN - Fe^{3+}$ units, whereas darker regions reflect more reduced environments enriched in $Fe^{2+} - CN - Fe^{2+}$ uncoordinated species. The contrast demonstrates a heterogeneous mixed-valence landscape within the film that evolves under electrical polarization.

To directly compare regions that did and did not experience voltage stress, Fig. 7c shows spectra acquired sequentially along the line scan in Fig. 7a. The first spectrum (cyan) exhibits the mixed PW-like signature with contributions from $Fe^{2+} - CN - Fe^{2+}$ uncoordinated species and a weaker $Fe^{2+} - CN - Fe^{3+}$ shoulder. At the electrically biased position (blue), the $\sim$ 2156 $cm^{-1}$ band dominates, reflecting oxidation and increased PB-like character. A subsequent spectrum recorded further along the scan (violet) resembles the initial state, indicating that the redox modulation is reversible and that the film recovers a similar mixed-valence distribution after removal of the voltage.

Fig. 7d shows a 2D Raman contour map that captures the spatial evolution of the vibrational response, clearly showing the $PW \rightarrow PB \rightarrow PW$ transition. The confined electric field beneath the C-AFM tip induces localized, reversible intercala tion and redistribution of interstitial K+ ions, which modulates the $Fe^{2+}/Fe^{3+}$ balance within the $Fe - CN - Fe$ network while preserving the structural integrity of the PBA lattice. The close correspondence between the field-induced Raman evolution and established vibrational fingerprints of PBAs with varying $K^+$ content and redox state provides strong evidence that nanoscale resistive switching in PW arises from localized $K^+$ migration coupled to Fe-center redox reconfiguration, rather than from irreversible chemical or structural changes or damages.

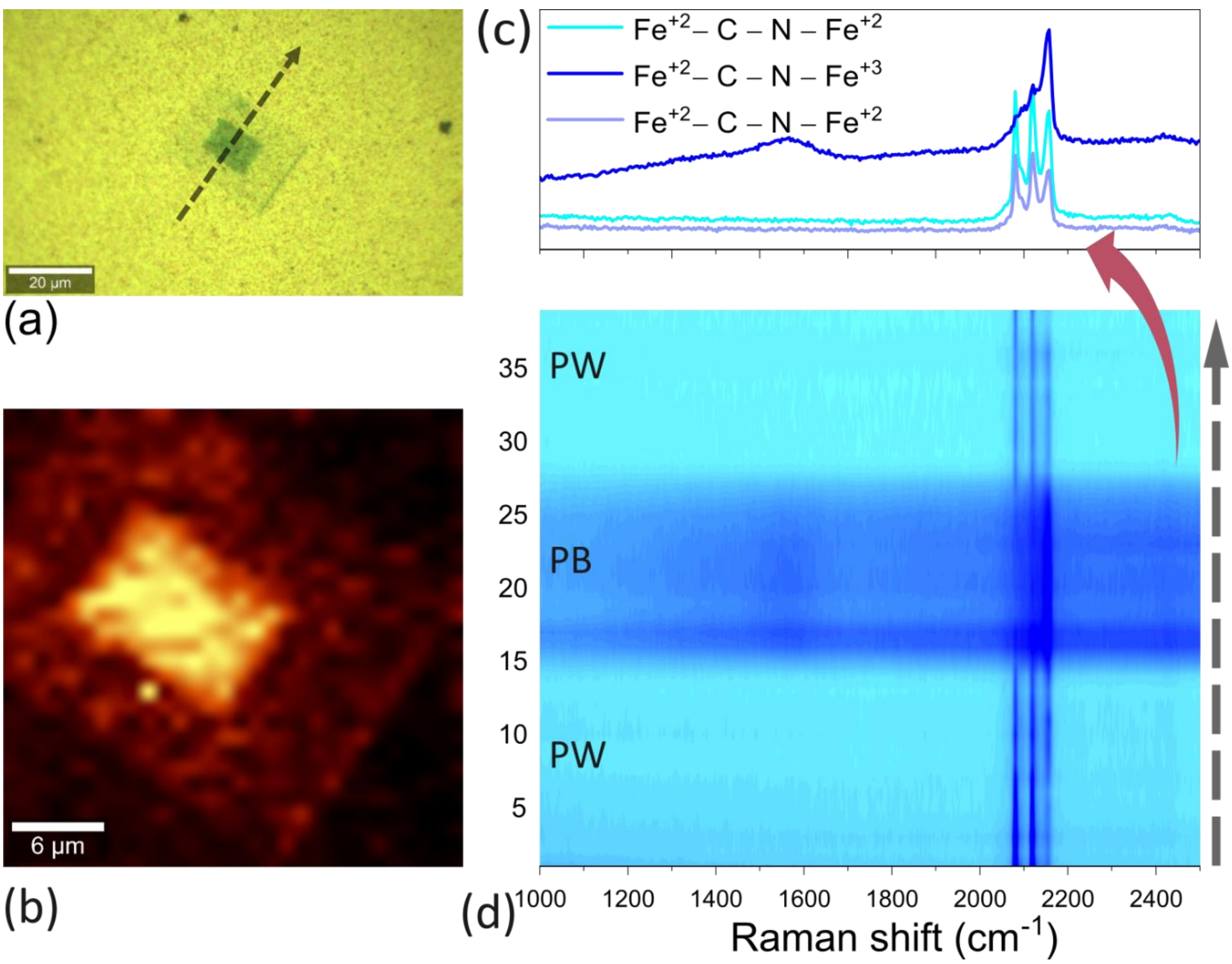

Figure 7. Influence of the C-AFM tip on the redox state of PW. (a) Optical micrograph of the $5 \times 5\ \mu m^2$ region in which the C-AFM experiment was performed 50 spots full Raman spectrum showing the transition from PW to PB and back to PW. The arow indicate direction the direction in which the Raman mapping was performed. (b) 2D confocal Raman intensity map acquired around the $\sim\ 2156\ cm^{-1}$. (c) Raman spectra highlighting the characteristic vibrational modes of PW ($2080, 2118,$ and $2156\ cm^{-1}$, shown in cyan/malibu) and PB (2122 and 2155 $cm^{-1}$, shown in blue). (d) 2D Raman contour map illustrating the temporal evolution of the Raman signal from 1000 to 2500 wavelength, and the distinct transition from PW and PB regions. The arow indicate Raman mapping direction.

## DISCUSSION

At the fundamental level, the resistive switching behavior observed in both PW and PB arises from a strongly localized ion intercalation electron coupling mechanism operating within the PBA lattice. The C-AFM data, finite-element simulations, and in situ Raman collectively indicate that switching is governed by subtle but highly effective short-range displacements of interstitial $K^+$ ions, which directly modulate the $Fe^{2+}/Fe^{3+}$ balance in the $Fe - C \equiv N - Fe$ network. Because the electronic conductivity in PBAs is mediated by intervalence electron hopping between adjacent $Fe^{2+}$ and $Fe^{3+}$ centers, even small, nanometric shifts in $K^+$ position are sufficient to alter local redox environments and, consequently, the connectivity of conductive pathways. In PW, where the $K^+$ content is higher, local $K^+$ rearrangement more readily stabilizes $Fe^{2+}$-rich regions, producing unipolar switching. In PB, the lower alkali-ion concentration leads to a more polarity-sensitive redistribution and thus bipolar switching.

Finite-element simulations show that the tip-induced electric field is confined to a $60\ nm$ volume immediately beneath the contact, meaning that the ion–electron coupling occurs within a tightly restricted nanoscale region. Because $K^+$ ions can only undergo limited short-range displacements within their coordination cavities, their screening effect remains local: they partially neutralize the applied electric field but only over nanometer-scale distances. This short-range ionic shielding prevents lateral field propagation and ensures that the switching process remains spatially isolated. As a direct consequence, each C-AFM measurement point behaves as an independent memristor element, and no interference is observed between neighboring sites, even at $100\ nm$ spacing, the smallest array pitch used in this study.

Another notable result of this study is the speed at which intercalation-induced commutation occurs. The literature shows that ultrafast resistive commutation sweep is observed almost exclusively in filamentary or phase-change devices, where atomic rearrangements or electronic transitions dominate [53–59]. Intercalation-based memristors, on the other hand, have historically been limited to voltage sweep rates below $1 - 10\ V/s$, constrained by the time required for long-range ionic migration and redox equilibrium. In this context, the performance of our PBA films is unprecedented: PW maintains clear resistive switching up to $200V/s$ and PB up to $50V/s$, maintaining well-defined SET and RESET transitions at rates one to two orders of magnitude faster than those previously reported for $Li$- or $Na$-based intercalation memristors [37,60,61].

Recent studies have highlighted the critical role of ion intercalation in governing resistive switching phenomena, challenging the long-standing filamentary paradigm tradi tionally associated with metal–insulator–metal memristive devices. In particular, Yueyue He and co-workers demonstrated that the insulating-to-conducting transition in printed $Ag/PBA/ITO$ memristors originates from the reversible extraction and insertion of $Na^+$ ions, rather than from the formation and rupture of metallic Ag filaments [32]. Through a combination of systematic electrical measurements and simulation studies, their work established ion-driven redox processes as the dominant mechanism controlling electronic transport. These findings are directly related to the results presented in this work. Based on this understanding, we report, for the first time, the nanoscale demonstration of resistive switching driven by high-rate K-ion intercalation, enabled by the open three-dimensional architecture and the intrinsically redoxactive $Fe - C \equiv N - Fe$ structure of PBAs, which favor fast and reversible coupling between ion transport and electronic conduction.

The in-situ Raman observations reinforce this mechanistic picture. The reversible evolution from PW-like to PB-like vi brational signatures and back demonstrates that the tip-driven electric stimulus induces transient $Fe^{2+}/Fe^{3+}$ rebalancing rather than irreversible structural modifications. This redox flexibility is a hallmark of the PBA framework and strongly supports the conclusion that nanoscale resistive switching originates from a reversible, local $K^+$ redistribution process. We also demonstrate that applying an electrical potential to the PW surface locally converts the material into a PB-like state enriched in $K^+$. This finding indicates that samples initially synthesized as PW can be electrically transformed in situ into a $K^+$-rich PB phase, which subsequently exhibits stable resistive switching behavior and can be operated in this modified redox state. Taken together, these findings establish that the RS behavior in PBAs is inherently material-governed, highly localized, and reproducible, dictated by short-range ionic motion and the intrinsic three-dimensional connectivity of the PBA lattice. Confocal Raman mapping was performed to resolve the spatial distribution of redox states in the PW and PB films. Because the two phases exhibit distinct $\nu(C \equiv N)$ stretching frequencies PW (K-rich, dominated) in the $2080 - 2120 cm^{-1}$ range and PB ( dominated) in the $2120 - 2160\ cm^{-1}$ range— Raman imaging provides a direct chemical contrast between them. In our measurements, regions with high intensity near $\sim 2156\ cm^{-1}$ appear as orange areas in the confocal map, corresponding to PB-rich domains, whereas darker regions denote low intensity associated with PW-rich areas. This spatial variation confirms the coexistence of locally oxidized (PB-like) and reduced (PW-like) environments across the film surface.

The CN stretching mode is highly sensitive to $K^+$ occupancy in the PBA framework; increased $K^+$ stabilizes $Fe^{2+}$ and generates PW-like spectral features, while reduced $K^+$ content shifts the spectrum toward PB-like peaks. Accordingly, the Raman maps also provide insight into the distribution of interstitial $K^+$ ions and reveal the degree of homogeneity in their incorporation.

Furthermore, in situ Raman spectra acquired at sites previously biased by the C-AFM tip show a reversible trans formation between the PW and PB signatures, confirming a localized $PW \rightarrow PB \rightarrow PW$ redox cycle induced by the applied electric field. These results demonstrate that K-ion redistribution occurs within a confined nanoscale region beneath the tip and directly modulates the local $Fe^{2+}/Fe^{3+}$ balance. These spectral changes directly confirm bias-induced redox switching and localized K-ion movement beneath the C-AFM tip, providing strong support for the ion-intercalation mechanism underlying the observed memristive behavior.

The combination of ionic confinement, reversible redox chemistry, and the tridimensional, open-framework structure gives PBAs a distinct advantage for nanoscale device integration. Because the switching volume is inherently limited by the host lattice geometry and does not rely on long-range ion transport, individual memristive sites can be patterned at $\leq\ 100\ nm$ spacing without electrical interference, a key requirement for high-density crossbar architectures. Thus, K-ion intercalation in Prussian blue analogs constitutes a fundamentally different resistive switching mechanism from conventional filamentary or phase-change processes. The strong nanoscale confinement of the switching region enables reliable downscaling to dimensions required for next-gener ation integrated architectures, while avoiding irreversible structural damage. Combined with single-step, aqueous, roomtemperature electrodeposition and full CMOS compatibility, PBAs offer a scalable, low-cost materials platform with clear advantages for high-density memory technologies.

## CONCLUSIONS

In this work, we demonstrate that K-ion intercalation acts as the fundamental switching mechanism in the memristive PBA framework and can be directly resolved at the nanoscale. Using C-AFM, we electrically control and spatially localize reversible conductance changes within sub-100 nm volumes, correlating $K^+$ redistribution with $Fe^{2+}/Fe^{3+}$ redox modulation in the PBA lattice. Supported by finite-element simulations and in situ Raman spectroscopy, our results show that resistive switching in PW and PB originates from short-range, reversible K-ion motion that modulates intervalence electron hopping, without filament formation or permanent structural changes. The switching behavior is intrinsically polarity-selective and governed by the initial redox state of each phase: oxidation-driven in PW and reduction-driven in PB. These processes are mediated by small-polaron transport and strongly influenced by ionic mobility, leading to distinct switching kinetics. Accordingly, PW sustains switching at scan rates up to $200\ V/s$, whereas PB is kinetically limited to $50\ V/s$, highlighting the critical role of K-ion concentration in enabling fast redox dynamics.

The nanoscale confinement of the electric field enables spatially isolated switching events, supporting device integration with lateral dimensions below 100 nm without interference between them. Combined with a single-step, room-temperature aqueous fabrication process, these results establish PBAs as a chemically versatile and earth-abundant platform for high-rate, scalable memristive devices. More broadly, this work translates a well-established battery intercalation mechanism into nanoscale memristive operation, positioning electrodeposited Prussian blue analogs as a new class of fast intercalation-based memristors.

**Supporting Information**

Supporting Information is available from the ACS Publications or from the author.

**Acknowledgements**

Research supported by the project PID2022-139586NB-C44, funded by MCIN/AEI/10.13039/501100011033 and FEDER, EU; and supported by the Ramón y Cajal grant (RYC2022-035618- I), funded by MCIU/AEI/10.13039/501100011033, and by the FSE+. F.A.A. is a Research Fellow of the F.R.S.-FNRS. C.K.M. thanks the Deutsche Forschungsgemeinschaft for funding (No. 531524052). The authors acknowledge funding from the EU (Pathfinder-4D-NMR 101099676), the Spanish MCIN (Unit of Excellence “Maria de Maeztu” CEX2019-000919-M). We are grateful for the funding provided by the Perte Chip Chairs of the Ministerio para la Transformación Digital y de la Función Pública, the European Union – NextGenerationEU. R.T-C. thanks the University of Valenciana for the Banc Santander grant (Santander UV25-23) and “la Caixa” Foundation (Neuron2Dblue).

**Conflict of Interest**

The authors declare no conflict of interest

**Data Availability Statement**

The data that supports the findings of this study are available in the Supporting Information of this article and at the GitHub.

METHODS

## PBAs Synthesis

The dielectric layer, consisting of a Prussian blue analog (PBA) thin film, was deposited at room temperature by electrochemical deposition in potentiostatic mode using an electrochemical workstation (Ivium CompactStat, Eind hoven, Netherlands). Electrodeposition was carried out in a conventional three-electrode configuration controlled by a potentiostat. The working electrode (WE) was an Au/Cr/Si substrate, fabricated by electron-beam evaporation of a 5 nm chromium adhesion layer followed by a 50 nm gold layer onto a (100)-oriented silicon wafer (1 × 1 cm2), under a base pressure of 10-5 Pa. Film growth was confined to a circular area of 0.5cm2 defined by an adhesive tape mask applied during sample preparation. A platinum foil was used as the counter electrode (CE), and a saturated calomel electrode (SCE) served as the reference electrode (RE). The electrolyte consisted of an aqueous solution containing 1.0 M KCl, 5.0 mM HCl, 0.5 mM FeCl3, and 0.5 mM K3Fe(CN)6 (ACS grade, > 99%, Sigma-Aldrich, Darmstadt, Germany), prepared in 100 mL of deionized water with the pH adjusted to 2. Film deposition was achieved by applying a constant potential versus the SCE, enabling precise control of the electrodepo sition process by limiting the total transferred charge. For all samples, a total charge of 10 mC was deposited, yielding estimated film thicknesses of approximately 300 nm for Prussian White (PW) films deposited at 0.1 V and 500 nm for Prussian Blue (PB) films deposited at 0.3 V. All depositions were performed at 25 °C. Notably, no post-deposition thermal or chemical treatments were required, which simplifies the fabrication process and reduces overall processing complexity. Additional details on the electrochemical setup and deposition protocol can be found in [39].

### Morphology and composition

Morphological and compositional properties were analyzed by field emission scanning electron microscopy (FEG-SEM, TESCAN CLARA, Brno, Czech Republic) equipped with an energy-dispersive X-ray (EDX) detector (Ultim Max 65 SDD, Oxford Instruments, Wiesbaden, Germany) at 20 keV and a Raman system (Witec RISE, Ulm, Germany). Raman measure ments were performed with a 532 nm laser at 0.4 mW.

### C-AFM measurements

Conductive Atomic Force Microscopy measurements are conducted on a Bruker Icon Dimension equipped with a CAFM module. The measurements are obtained with SCM-PITV2 probes from Bruker. Those have a Pt/Ir metallic coating and a25-nm radius, an elastic constant of3 Nm-1 and a resonance frequency of 75 kHz. A thermal tune calibration of the tip is done before the measurements to update the elastic constant value. During the C-AFM scans, the deflection setpoint is kept between 50 and 70 nm and adjusted to optimize the signal quality and stability. I-V curves are then acquired at specific locations with a volage sweep between 0 and 10 V on every sample while keeping the same parameters as for the scan. The data treatment was carried out using Gwyddion and Nanoscope softwares, as well as the pySPM library [Scholder, O. scholi/pyspm: pyspm v0.2.16 (version v0.2.16). https://doi. org/10.5281/zenodo.998575].

### Simulation

Finite-element simulations were carried out in COMSOL Multiphysics using an axisymmetric geometry. The Solid Mechanics and Electric Currents modules were employed sequentially to first determine the mechanical deformation induced by the CAFM tip and subsequently compute the resulting electric potential and current distribution. During the mechanical step, a normal load of 180 nN, corresponding to the experimental CAFM conditions was applied to the tip. The tip was modeled as a rigid indenter penetrating a linear-elastic, isotropic medium representing the sample. The resulting indentation profile and stress distribution within the material were obtained from this step and used as input for electrical simulation. For the electrical step, a bias of 1 V was applied to the tip while the bottom electrode was grounded. The lateral boundaries of the sample were set as electrically insulating so that no current could exit through the sidewalls.

The electric potential distribution and total current through the bottom electrode were then computed assuming an isotropic conductivity. Two sample thicknesses were considered: 500 nm for PB and 300 nm for PW. The lateral size of the computational domain and the mesh densities were chosen based on convergence analyses performed on (i) the indentation depth, (ii) the total transmitted current, (iii) the potential evaluated 3 nm below the surface, and (iv) the residual current leaking through the lateral boundaries. These criteria ensured that both the mechanical deformation and the electrical response were independent of the domain size and mesh discretization.

**Data analysis and plots.**

All figures presented in this work were generated using a custom method, Python-based plotting framework that we developed for this study. To ensure full transparency and reproducibility, the complete repository including the raw data, processing scripts, and figure-generation code, is openly available on GitHub (https://github.com/AlTemporao/k-ionintercalation-memristors-in-pbas). No numerical simulations or model fitting were performed; all results reflect direct analysis of the experimental data. Based on the roughness measured by the AFM, we correlated the thickness variations of the PW (and PB) layers at different locations with the IV curves measured at those locations. Then, using in-house Python scripts, we plotted those IV curves at the LRS with a color scale to identify thickness differences between points. We used the LRS as a reference state for this figure because if the RS process depends on the thickness of the switching layer, a clear correlation between the current and the thickness variations is expected.